\documentclass[aps,prb,twocolumn,superscriptaddress,floatfix,english,amsmath,amssymmb,longbibliography,10pt]{revtex4-2}
\usepackage{graphicx}
\usepackage{dcolumn}  
\usepackage{bm}       
\usepackage{hyperref} 
\usepackage{color}
\usepackage[T1]{fontenc}
\usepackage{amsmath}
\usepackage{amssymb}
\usepackage{esint}
\usepackage{comment}

\usepackage{amsbsy}
\usepackage{babel}
\usepackage{enumerate}
\usepackage{epsf}
\usepackage{esint}
\usepackage{epsfig,psfrag}
\usepackage{float}
\usepackage{latexsym}
\usepackage{mathbbol}
\usepackage[usenames,dvipsnames]{xcolor}
\usepackage[mathscr]{eucal}
\usepackage{wrapfig} 
\usepackage{ulem}

\begin{abstract}
	We study a cylindrical plasmonic waveguide consisting of a magnetic Weyl semimetal embedded in a dielectric medium. We determine the dispersion relation of the surface plasmon polaritons and show how it depends on the plasma frequency, the radius of the semimetal and the separation between the nodes. 
	We show that the band structure, which modifies the electrodynamics in the medium, manifests itself through a pronounced asymmetry in the dispersion curves and a giant splitting in the group velocity, with the orbital angular momentum 
	as a control parameter for the direction of propagation.
\end{abstract}

\begin{document}
	\title{Nonreciprocal Weyl semimetal waveguide}
	\author{Marco Peluso}
	\affiliation{Dipartimento di Scienza Applicata e Tecnologia, Politecnico di Torino, Corso Duca degli Abruzzi 24, 10129, Torino, Italy }
	\author{Alessandro De Martino}
	\affiliation{Department of Mathematics, City St~George's, University of London,
		Northampton Square, EC1V OHB London, United Kingdom}
	\author{Reinhold Egger}
	\affiliation{Institut f\"ur Theoretische Physik,
		Heinrich-Heine-Universit\"at, D-40225  D\"usseldorf, Germany}
	\author{Francesco Buccheri}
	\affiliation{Dipartimento di Scienza Applicata e Tecnologia, Politecnico di Torino, Corso Duca degli Abruzzi 24, 10129, Torino, Italy }
	\affiliation{INFN Sezione di Torino, Via P. Giuria 1, 10125, Torino, Italy}
	
	\date{\today}
	\maketitle
	
	\section{Introduction}
	Surface plasmon polaritons (SPPs) are coherent excitations of electrons and radiation, confined at the interface between a metal and a dielectric. The possibility of confining the radiation on sub-wavelength scales allows to greatly enhance the intensity of the electromagnetic field and its interaction with matter \cite{Yu2019}. A plasmonic waveguide, in particular, exploits the SPPs to overcome the limitations in light confinement caused by diffraction and to reduce the size of the device, which can reach diameters down to tens of nm \cite{Oulton2009,Gramotnev2010}. It is also the basic component of a number of devices, e.g., SPP nanolasers \cite{Oulton2009,Hill2009}.
	Controlling the direction of SPPs is often necessary to reduce noise and desirable to various applications, e.g., circulators or amplifiers \cite{Camley1987}. It requires, however, specially engineered circuit elements \cite{Lin2013,Liu2012} or interfaces \cite{Huang2013}. 
	
	In this work, we explore the possibility of using a Weyl semimetal (WS) as constituent element of the waveguide. WSs are topological materials, exhibiting non-degenerate band touching points in an otherwise gapped Brillouin zone \cite{Armitage2018}. Since their discovery \cite{Xu2015,Lv2015}, a large number of compounds have been shown to exhibit a WS phase \cite{Kruthoff2017,Bradlyn2017,Vergniory2019}, eliciting intense theoretical and experimental interest. These materials possess nontrivial transport properties \cite{Son2013,Pellegrino2015}, which can be traced back to the presence of an axionic term in the emergent electrodynamics \cite{Wilczek1987,Sekine2021}, directly connected to a chiral anomaly \cite{Fujikawa}. 
	Our approach exploits these features to propose a novel way to control the propagation of plasmonic excitations.
	More in detail, the axionic term in a WS is quasi-universal, in that it is fully determined by universal constants, while the realization-specific structure of the electronic band only enters via the separation of the band-touching points. This term encodes the anomalous Hall effect, the chiral magnetic response \cite{Zyuzin2012,Vazifeh2013,Chen2013}, and the optical activity \cite{Kargarian2015,Chen2019}. The giant nonreciprocity in magnetic WSs results in significant magneto-optical effects, which can be exploited to design plasmonic circuit elements, e.g., sub-wavelength optical insulators \cite{Asadchy2020}. 
	
	At planar interfaces between WSs and dielectrics, unique properties of the SPPs emerge from the strong coupling between light 
	and the chiral surface Fermi arcs \cite{Song2017,Andolina2018}. As a consequence, SPPs exhibit an anisotropic dispersion in a half-space geometry \cite{Zhou2015,Hofmann2016,Kotov2018}. Analogously, the optical response of thin films \cite{Tamaya2019}, hybrid layered structures \cite{Oskoui2019,Oskoui2023}, and magnetic domain interfaces \cite{Zyuzin2015,Lu2021} exhibits a strongly anisotropic character \cite{Guo2023}. 
	Despite its potential technological impact, a WS waveguide with compact transverse section has not yet been studied. Importantly, a compact section allows a previously overlooked effect, namely, the nonreciprocity of light propagation with respect to the orbital angular momentum (OAM) quantum number.
	
	\begin{figure}
		\centering
		\includegraphics[width=0.75\linewidth]{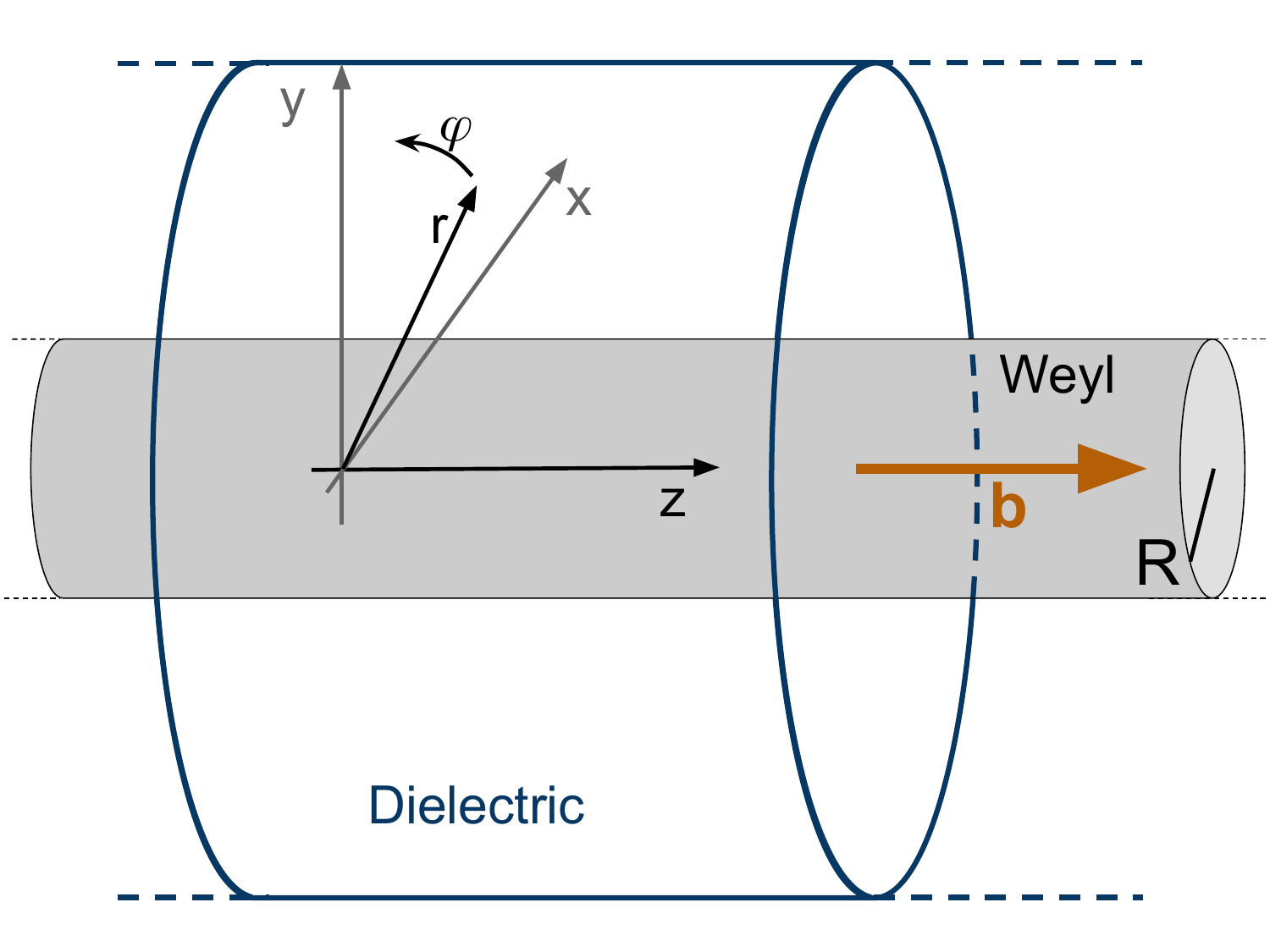}
		\caption{Cylindrical waveguide scheme: a magnetic WS core surrounded by a dielectric medium. The wavevector $\bf b$ along the $z$ direction describes the Weyl node separation in momentum space.}
		\label{fig:Weylguide}
	\end{figure}
	
	To show this, we investigate a one-dimensional waveguide with circular section, in which a dielectric coating surrounds a topological magnetic WS cylindrical wire,  with the magnetization along the axis.
	This design exploits the fact that only a portion of the energy is carried in the dissipative medium, so that long-range SPPs are supported. In addition, the reduced density of states and the electronic band structure enhance the propagation of the electromagnetic field in this class of materials \cite{Sukhachov2022}.
	Light beams can be characterized by a polarization, or by an OAM index \cite{Allen1992}. Such Laguerre-Gaussian beams can be generated with a high degree of control \cite{Matsumoto2008}.
	In this work, we show that the interplay between the transverse confinement in the WS wire and the axionic term determines a novel nonreciprocity of the dispersion in the OAM. 
	Remarkably, plasmons with OAM equal in modulus but opposite in sign propagate with a different, and in some regimes opposite, group velocity. This previously overlooked phenomenon can be exploited to control the signal propagation within the Weyl semimetal plasmonic waveguide (WPW), providing an additional degree of freedom to the bands available for energy and information transport \cite{Akimov2007,Krasavin2010}.

	\section{Model}
	We consider a magnetic WS with a single pair of band crossings, or Weyl nodes. Such a phase has been predicted in various materials, e.g., $\mbox{Eu}\mbox{Cd}_{2}\mbox{As}_{2}$ \cite{Wang2019,Krishna2018,Wang2016}, $\mbox{Hg}\mbox{Cr}_{2}\mbox{Se}_{4}$ \cite{Xu2011}, $\mbox{Mn}\mbox{Bi}_{2}\mbox{Te}_{4}$ \cite{Li2019}, $\mbox{Mn}\mbox{Sn}_{2}\mbox{Sb}_{2}\mbox{Te}_{6}$ \cite{Gao2023,Boulton2024}, $\mbox{K}_{2}\mbox{Mn}_{3}\left(\mbox{As}\mbox{O}_{4}\right)_{3}$, $\mbox{X}\mbox{Cr}\mbox{Te}$, (X=K, Rb) \cite{Liu2024}, $\mbox{Eu}_{2}\mbox{Ir}_{2}\mbox{O}_{7}$ \cite{Sushkov2015}, and manufacturing of Weyl semimetal nanowires is within present-day capabilities \cite{cheon2025}.
	The universal low-energy Hamiltonian describing the electron dynamics 
	in the vicinity of a Weyl node at ${\mathbf k}=\pm \mathbf{b}$, is given by
	\begin{equation}
		H_\chi(\mathbf{k})=\hbar v_F \bm{\sigma} \cdot \left( {\mathbf k} - \chi {\mathbf b}\right)\;, 
		\quad \chi =\pm, 
	\end{equation} 
	where $\bm{\sigma}=(\sigma^x,\sigma^y,\sigma^z)$ are the Pauli matrices, $\mathbf{k}$ the electronic crystal momentum, and $v_F$ is the Fermi velocity, typically of order $\sim 10^5\mbox{m}/\mbox{s}$. We denote the vector that separates the Weyl nodes as $2\mathbf{b}$ and set its orientation along the $k_{z}$ axis. A cylinder of radius $R$, with its axis along the $z$ direction, is considered, see Fig. \ref{fig:Weylguide}. 
	The semimetal is surrounded by a cylindrical dielectric, with relative permittivity $\epsilon_d$ approximately independent of 
	radiation frequency and wavevector.
	As we are interested in modes whose intensity decays radially in the dielectric, we assume that the outer diameter of the coating is much larger than all the other length scales.
	
	The electronic states in this configuration have been studied in \cite{Bardarson2019,Sukhachov2020,DeMartino2021,Liu2023}, but we will describe the SPPs via a different approach, using the dynamics of the electromagnetic field. When coupled to the Weyl electrons, the dynamics can be reformulated in terms of a Dirac action; in doing so, the electromagnetic part of the action acquires an additional term, proportional to the chiral anomaly \cite{Grushin2012,Zyuzin2012,Vazifeh2013,Chen2013,Burkov2018}.
	Such {$\theta$} term is universal, i.e., independent of the realization of the WS phase up to the separation between the Weyl nodes in the Brillouin zone, and so are its observable consequences, e.g., the anomalous Hall and the chiral magnetic effects.
	It follows that, neglecting the possible shift in energy between the Weyl nodes, the classical dynamics of the electromagnetic field is governed by the axion electrodynamics equations \cite{Wilczek1987,Armitage2019}
	\begin{align}
		\nabla\cdot\mathbf{E}	& =	\frac{\rho_{e}}  
		{\varepsilon_{0}}+\frac{2\alpha c}{\pi}\mathbf{b}\cdot\mathbf{B} \,, \label{del.E} \\
		\nabla\times\mathbf{E}	& =	-\partial_{t}\mathbf{B} \,, \label{delxE} \\
		\nabla\cdot\mathbf{B}	&=	0 \,, \label{del.B}\\
		\nabla\times\mathbf{B}	&=	\frac{1}{c^{2}}\partial_{t}\mathbf{E}+
		\mu_{0}\mathbf{j}_e-\frac{2\alpha }{\pi c}\mathbf{b}\times\mathbf{E}\, .  \label{delxB}
	\end{align}
	Here \mbox{$\alpha={e^{2}}/{4 \pi \varepsilon_{0}\hbar c}\approx {1}/{137}$}
	is the fine structure constant, $\rho_{e}$ and $\mathbf{j}_e$ are the total charge and current densities, 
	while $\varepsilon_{0}$ is the permittivity of the vacuum.
	While the homogeneous equations \eqref{delxE} and \eqref{del.B} are unaltered, 
	the Weyl node separation $2\bf b$
	explicitly appears in the anomalous density in \eqref{del.E} and in the Amp\`ere-Maxwell 
	law \eqref{delxB}. These anomalous terms break time reversal invariance of Maxwell's equations.
	Throughout this paper, we consider a monochromatic mode of angular frequency $\omega$, 
	$\mathbf{E}(t,{\bf r}) = e^{-i\omega t} \mathbf{E}({\bf r})$, 
	where $\mathbf{E}({\bf r})$ is the complex field amplitude, and we omit the frequency argument.
	
	The interaction of light with the electrons in the semimetal elicits a current density 
	$\mathbf{j}_e=\left[\sigma-i\omega \varepsilon_0(\epsilon_W-1)\right]\mathbf{E}$. 
	Here, $\sigma(\omega)$ is the dynamical conductivity of a Dirac semimetal in the long wavelength limit and  
	$\epsilon_{W}$ is the static background relative dielectric constant.
	While the conductivity $\sigma$ is diagonal, the vector $\mathbf{b}$ generates the off-diagonal terms, see \eqref{delxB}, which produce the quasi-universal anomalous Hall effect \cite{Burkov2014,Zyuzin2012}. 
	The electronic matter determines the functional form of the relative permittivity $\mathcal{E}{\left(\omega\right)}=\epsilon_{W}+i{\sigma{\left(\omega\right)}}/{\varepsilon_0\omega}$ \cite{AshcroftMermin}.
	In the local-response and low-temperature approximations, it takes the form
	\cite{Kotov2016,Kotov2018}
	\begin{equation} \label{permittivity}
		\mathcal{E} = \epsilon_{W}\left( 1-\frac{\omega_p^2}{\omega^{2}} \right)
		+\frac{\omega_p^2}{4\omega_{F}^{2}}
		\left[\ln\frac{4\omega_{c}^{2}}{\left|\omega^{2}-4\omega_{F}^{2}\right|}
		+i\pi\Theta\left(\omega-2\omega_{F}\right)\right]
		\;,
	\end{equation}
	where finite lifetime effects are neglected. Here \mbox{$\omega_p^2={e^2\omega_F^2}/{3\pi^2 \epsilon_W \varepsilon_0 \hbar v_F}$}
	denotes the squared plasma frequency in the Drude-like (or single-band) approximation, 
	in which one retains only the first term in Eq.~\eqref{permittivity}. 
	\mbox{$E_F=\hbar\omega_F$} is the Fermi energy and $E_c=\hbar\omega_c$ is a cutoff energy, 
	determined by the range of energies in which a linear behavior 
	is a good description of the electronic spectrum \cite{Kargarian2015} 
	and transitions to other bands can be excluded. 
	While the plasma frequency is properly defined as the zero of the real part of \eqref{permittivity},
	$\omega_p$ is a very good approximation for realistic values of the cutoff and Fermi energies. 
	Interestingly, $\omega_p$ can be comparable to $\omega_{F}$.
	
	There are three relevant frequency scales. The plasma frequency $\omega_p$ and the frequency $cb$ associated to the separation between the Weyl nodes are characteristic of the bulk. 
	The finite section introduces the scale $c/R$, associated to the transverse size.
	The dispersions of the electromagnetic field eigenmodes depend on the two dimensionless combinations
	\begin{equation}
		\rho=\frac{\omega_p R}{c},\qquad \beta=\frac{\alpha c b}{\pi\omega_p},
	\end{equation}
	which parametrize the radius of the inner cylinder of the WPW and the separation between the Weyl nodes.

	\section{SPP modes}
	We now proceed to calculate the normal modes of the modified Maxwell electrodynamics. 
	We first solve the wave equation in the bulk; then, we determine the dispersion relation of the SPPs by imposing the appropriate matching conditions 
	for the fields at the interface between the WS and the dielectric. We focus throughout on modes 
	decaying on both sides of the interface.
	
	The solution of equations \eqref{del.E}-\eqref{delxB} in the frequency domain exploits 
	the cylindrical symmetry of the system. In cylindrical coordinates $(r,\varphi,z)$, one writes the electric field in the form
	\begin{equation}\label{eq:fieldform}
		\mathbf{E}(r,\varphi,z) = \sum_m e^{iq_zz+im\varphi}\mathbf{E}_m(r)\;,
	\end{equation}
	where $q_z$ is the wavevector along the axis, $m\in\mathbb{Z}$ the OAM label.
	The vector field $\mathbf{E}_m(r)$ is further constrained by the cylindrical symmetry, see Appendix \ref{app:solution}.
	There, we show that the dispersion relation is determined by seeking a solution in the form \eqref{eq:fieldform} 
	and subsequently imposing the consistency of the axion electrodynamics equations. 
	The presence of the extra term in the non-homogeneous Maxwell equations modifies 
	the dispersion relation of the electromagnetic modes in the bulk of the material. 
	The latter is implicitly determined by the equation
	\begin{equation}\label{secular}
		\left(\omega^2\mathcal{E}-c^2q^2\right)^2\mathcal{E}
		=\left(\frac{2\alpha c b}{\pi}\right)^2 \left(\omega^2\mathcal{E}-c^2q_\perp^2\right)\;,
	\end{equation} 
	where 
	$q=\sqrt{q_z^2+q_\perp^2}$, and  $q_\perp=\sqrt{q_x^2+q_y^2}$ 
	is the modulus of the radial component of the wavevector. 
	
	In the presence of an interface at $r=R$, the wavevector in the radial direction is not a conserved quantity, but is instead determined from \eqref{secular} as a function of the axial momentum and the frequency.  This yields real solutions, which correspond to the waveguide modes, as well as imaginary solutions $q_\perp=-i\kappa$, which decay exponentially from the interface toward the axis and are associated with the SPP modes. These are the object of this work and are characterized by the inverse localization length
	\begin{equation}\label{lambda}
		\kappa_{\pm} = \sqrt{q_z^{2}-\frac{\omega^2}{c^2}\mathcal{E}+\frac{2}{\mathcal{E}}\left(\frac{\alpha b}{\pi}\right)^2\pm \frac{2\alpha b}{\pi}\sqrt{\frac{q_{z}^{2}}{\mathcal{E}}+\left(\frac{\alpha b}{\pi\mathcal{E}}\right)^2}}\;.
	\end{equation}
	\begin{figure}[h]
		\centering
		\includegraphics[width=\linewidth]{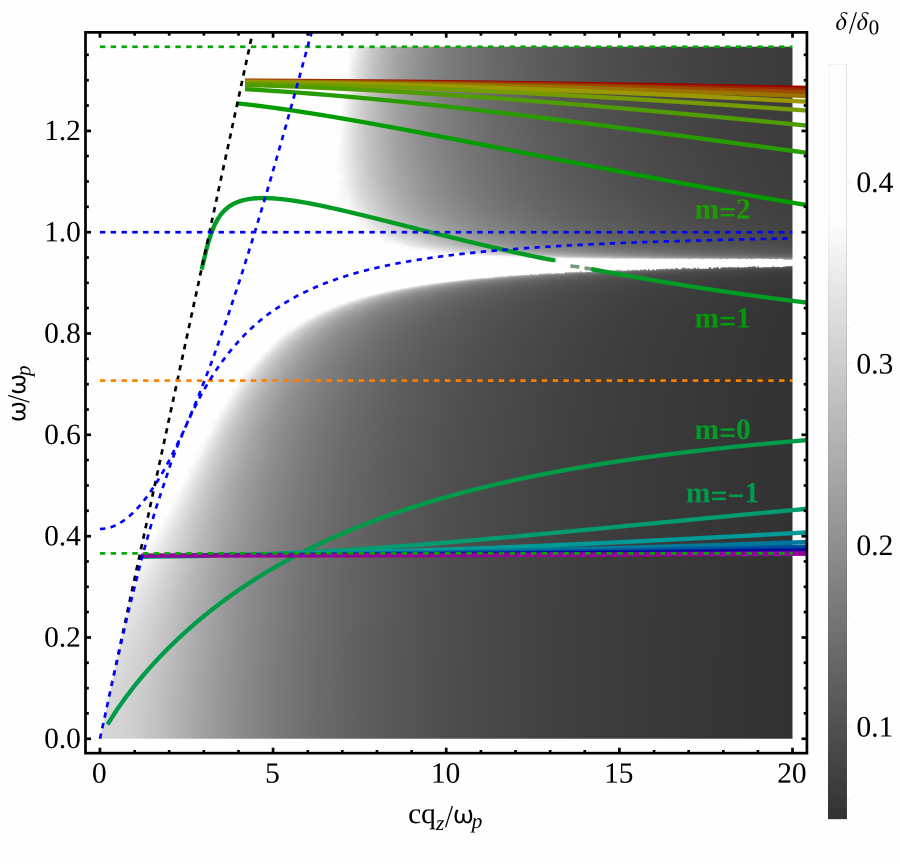}
		\caption{Dispersion relations of the SPP modes of the cylindrical waveguide at small wavevector. 
			For $\rho=0.1$, $\beta=10$, $\omega_c/\omega_p=10$, $\omega_F/\omega_p=1$, 
			the shown modes have OAM label between $m=-10$ (lowest purple line) and $m=10$ (highest red line), 
			with color ordering. Background: SPP penetration length in units of the metal skin depth $\delta/\delta_0$
			for $\epsilon_{W}=\epsilon_{d}=10$. The dashing in the dispersions visible in the $m=1$ line indicates a region where the modes do not have a surface component.
			The dashed guide lines represent analytical expressions, which we compare to the numerical solutions. We show the asymptotic lines \eqref{omegainfinity} (orange) 
			and \eqref{omegapm} (green). We also show in blue the boundary of the region with mixed surface and bulk modes \eqref{omegamix}, 
			as well as $\omega_p$ and the degeneracy line discussed below Eq. \eqref{omegamix}. 
			The formulae are more accurate away from $\omega_p$, but discrepancies always stay below $10\%$. 
			All plasmonic dispersions end on the dispersion of the light in the dielectric medium (black dashed), on whose left they are not localized.
			\label{fig:dispersionsWPW}}
	\end{figure}
	The interface delimits a compact section of material, so that the light modes are labeled by the OAM \cite{Allen1992}, rather than the wavevector perpendicular to the light propagation.
	In Fig. \ref{fig:dispersionsWPW} we show the dispersion relations for \mbox{$\rho=0.1$}, \mbox{$\beta=10$}. 
	In order to understand these curves, let us start by noting that, when $cq_z/\omega_p\to\infty$, 
	they all tend to the asymptotic value 
	\begin{equation}\label{omegainfinity}
		\frac{\omega_\infty}{\omega_p}=\sqrt{\frac{\epsilon_W}{\epsilon_d+\epsilon_W}}\;,
	\end{equation}
	formally the same as for conventional metals \cite{Grosso2000}. 
	However, in a WPW, a novel scaling regime appears, in which the longitudinal wavevector is coupled with the OAM
	\begin{equation}\label{largeqz}
		\frac{\omega_{m}}{\omega_{\infty}} \simeq	 1+\frac{m\beta\sqrt{\frac{\epsilon_{d}+\epsilon_{W}}{\epsilon_{W}}}-\frac{\epsilon_{d}}{2}}{\epsilon_{d}+\epsilon_{W}}\frac{1}{Rq_{z}} +\mathcal{O}\left(\frac{1}{R^2q_{z}^2}\right) \;,
	\end{equation}
	when  $Rq_{z} \gg 1,\;{m\beta\omega_p}/{\omega_{\infty}}$. 
	The asymmetry of the dispersion under $m\to-m$ in \eqref{largeqz}, quantifies the observed difference in curves and shows that there are modes with negative group velocity $v=d\omega/dq_z$. 
	Eq. \eqref{largeqz}, rigorously valid in the Drude approximation for the permittivity, is a good approximation for the full SPP dispersions as well, because typical SPP frequencies are around $\omega_\infty<\omega_p$, see Eq. \eqref{omegainfinity}.
	The limit $m\to\pm\infty$ at fixed value of $q_z$ identifies instead two distinct lines
	\begin{equation}\label{omegapm}
		\frac{\omega_\pm}{\omega_p} = \frac{\epsilon_W}{\sqrt{\epsilon_W\left(\epsilon_W+\epsilon_d\right)+\beta^2}\mp\beta}\;,
	\end{equation}
	which are accumulation points for positive and negative OAM. 
	While the dispersion with $m=0$ is always the lowest-frequency mode in a metallic waveguide, this is not the case in a WPW.
	The SPP dispersions are symmetric under $q_z\to-q_z$. Our solution correctly reproduces the planar limit $R\to\infty$ and the normal metal limit $b\to0$, known in implicit form \cite{Pfeiffer1974,Stratton}.
	
	\section{Light propagation in the WPW}
	The SPP penetration in the WS bulk differs from that of its metallic counterpart \cite{Deng2021}. Indeed, the inverse penetration lengths $\kappa_\pm$ in Eq.~\eqref{lambda} have a finite imaginary part.
	More in detail, our solution is a superposition of two waves, each expressed in terms of modified Bessel functions. These behave approximately as an exponential, with a penetration depth given by $\delta_\pm=1/\Re\left[\kappa_{\pm}\right]$. In Fig. \ref{fig:dispersionsWPW}, we compare the maximal penetration depth $\delta=\max_{\pm}\delta_\pm$ with the metal skin depth $\delta_0=c/\omega_p$. The shorter wavelengths, on the right, are more localized. On the left, the localization length diverges on the line identified by the condition $\kappa_-=0$, see Eq. \eqref{lambda}.
	Using the Drude form of the permittivity \eqref{permittivity}, one obtains the analytic curve
	\begin{equation}\label{omegamix}
		\frac{\omega_{mix}}{\omega_p}=\frac{1}{\epsilon_W} \left[\sqrt{
			\epsilon_W\left(\frac{c^2q^2_z}{\omega_p^2}+\epsilon_W\right)+\beta^2}-\beta\right]\;.
	\end{equation}
	shown in the figure as a dashed blue line, also in good agreement with the numerical results obtained using the full permittivity.
	Between this line and the light dispersion, the modes have mixed character, a superposition between SPPs (exponentially localized) and waveguide (oscillating) modes. Purely waveguide modes are present in the roughly triangular region at the center of the figure, delimited by the plasma frequency and the degeneracy line identified by the condition $\kappa_+=\kappa_-$. Outside of these regions, we have SPP modes.
	\begin{figure}
		\centering
		\includegraphics[width=\linewidth]{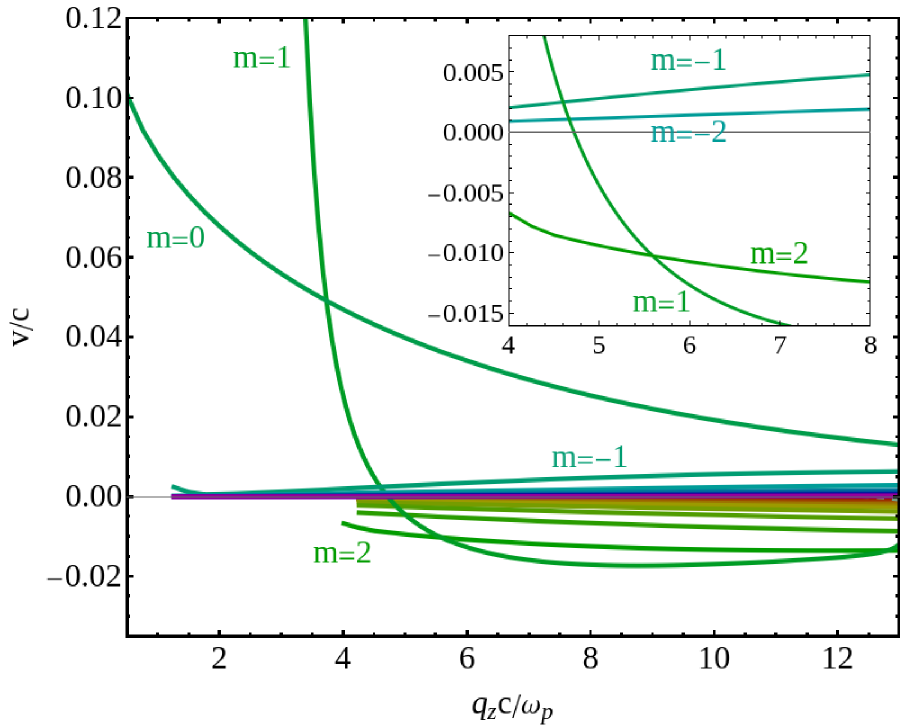}
		\caption{SPP group velocity for $\rho=0.1$, $\beta=10$, \mbox{$\omega_c/\omega_p=10$}, $\omega_F/\omega_p=1$, for the modes between $m=-10$ (purple) and $m=10$ (red), with color ordering (same color code as in Fig.~\ref{fig:dispersionsWPW}). Inset: zero of the $m=1$ mode and opposite velocities of the lowest OAM. The splitting due to the topological axion term is as large as the velocity itself and determines the change in sign.}
		\label{fig:velocityWPW}
	\end{figure}
	As can be seen in Fig. \ref{fig:dispersionsWPW}, our analytical expressions \eqref{omegainfinity}-\eqref{omegamix} 
	are accurate approximations of the asymptotic behaviors as long as their value is away from $\omega_p$. 
	The exact value can be obtained by solving the pertinent asymptotic conditions numerically, see Appendix \ref{app:bc}.
	We also observe from Fig. \ref{fig:dispersionsWPW} that the axionic term pushes down the frequency of the negative-$m$ modes. Compared to a metal, this also makes more SPP modes available with wavevectors.
	
	The group velocity of the SPPs is computed as the derivative of the numerical dispersion.
	The axionic term generates a giant nonreciprocity in the group velocities, controlled by the OAM $m$, which is the main result of this work:
	modes with opposite values of $m$ propagate with opposite group velocity in some regimes. 
	The proposed device does not require a magnetic field to control the wave propagation \cite{Camley1987}.
	We exemplify this observation in Fig. \ref{fig:velocityWPW}. Noticeably, the group velocities of the $m=\pm2$ modes have opposite sign. Moreover, the $m=1$ mode exhibits a zero and the associated sign change, shown in Fig. \ref{fig:velocityWPW} around $cq_z/\omega_p\approx 5$, which signals the onset of the short-wavelength regime, see Eq. \eqref{largeqz}.
	
	\section{Discussion}
	Our results show a strong effect due to the Weyl nodes on the SPP modes and a giant nonreciprocity under time-reversal, which inverts the OAM quantum number $m\to-m$.  While the occurrence of the splitting of the bands with opposite $m$ may be explained with the time-reversal symmetry breaking at a microscopic level by an intrinsic magnetization, in this class of materials, the effect on the propagation velocity appears to be giant, as large as the velocity itself. To understand this, we underline that SPPs arise from the hybridization of electromagnetic and electronic modes. The semiclassical dynamics of the electrons in WSs is largely influenced by the Berry curvature \cite{Son2013,Gangaraj2020}. As there is a net flux of Berry curvature in the region of the Brillouin zone between the Weyl nodes, one finds a nonzero expectation value of the electron angular momentum \cite{Schuler2020}. This is inherited by the radiation and determines a preferential sign of the OAM in the SPP dispersion \cite{Uenzelmann2020}.
	The origin of the effect lies in the anomalous term proportional to $\mathbf{b}$ in \eqref{delxB} and in its macroscopic manifestation, the anomalous Hall effect. This term enters the off-diagonal permittivity and is responsible for the gyrotropic responses. In order to quantify the role of the node splitting, we can directly compare the anomalous Hall responses of a well-studied non-topological ferromagnet, Fe \cite{Yao2004}, with that of a well-studied topological Weyl semimetal, $\mbox{Co}_3\mbox{Sn}_2\mbox{S}_2$ \cite{Liu2018}. Using the values provided in these references for the AHE, we expect an enhancement of at least $50\%$ in the latter material. We also point out that various electronic scattering mechanisms may destroy these surface plasmon features in ordinary ferromagnets. Conversely, the effect may be observable in Weyl semimetals, in which the band topology, the surface states and the associated anomalous Hall effect are robust against weak scattering, e.g., with phonons \cite{Buchhold2018,Buccheri2022}.
	The effect studied in this work qualitatively differs from previously noted asymmetries in SPP dispersions, as it does not require changing  the orientation of the incoming beam, nor the configuration of the guide \cite{Hentschel2017,Kotov2018,JalaliMola2019,Zhang2018,Shastri2021}. Conversely, a compact transverse size is essential. In infinite slab geometries, depending on the propagation direction, one would instead observe either reciprocal dispersion or opposite nonreciprocities on the opposite interface, leading to a cancellation of the net effect of the nodes. The SPP nonreciprocity qualitatively differs from the Faraday (Kerr) rotation \cite{Kargarian2015}, in that it relates the beam OAM and its propagation, as opposed to its polarization. 
	Our analysis exploits the long wavelength approximation of the dielectric function, which is valid for $\omega\ll\omega_F$ \cite{Zhou2015}. While we do not expect qualitative changes in the discussed regimes, inclusion of the $q$-dependence in the permittivity would allow to obtain 
	quantitatively more accurate dispersion curves and to explore larger wavevectors \cite{Zhou2015,Andolina2018}. 
	
	Our result does not rely on a specific material in which the Weyl semimetal phase is realized. A way of testing our predictions is to excite the different SPPs modes via Laguerre-Gaussian beams, which carry definite OAM \cite{Genevet2012,Kim2010,Knyazev2015} or illuminated metal tips \cite{Pellegrino2015}. Detection may exploit the strong coupling of the SPP modes to quantum dots \cite{Chang2006} or that the waveguide can filter out one sign of the OAM.
	Alternative platforms for realizing the SPP nonreciprocity, as well as the light-induced phenomena in OAM-filtered light, are a few among many interesting open questions.
	The proposed plasmon waveguide can be scaled down to nanometer size and operate down to THz frequencies while retaining the described efficient control mechanism. Such a device is central to many applications, including nanocircuits, holography, nanosensors, and nanophotonics \cite{Stockman2004,Gramotnev2010,Aharonovich2016,Chen2021,Jiang2024,Tredicucci2024}, may find use in classical communication between quantum architectures, and on-chip isolation \cite{Yuan2021}. The ability to control SSP propagation at the nanoscale holds significant promise for advancing both classical and quantum technologies.
	\subsubsection*{Data availability}
	The dataset needed to generate the figures \ref{fig:dispersionsWPW}, \ref{fig:velocityWPW}, \ref{fig:full} is avaliable at \cite{data}.
	
	\subsubsection*{Acknowledgments}
	We thank F. Dolcini, F. Rossi, E. Di Fabrizio and F. Medina Cuy for the interesting discussions and references. MP is founded through DM 118/2023 - Inv. 4.1, project "Light-matter interactions in topological semimetals", CUP E14D23001640006, Piano Nazionale di Ripresa e Resilienza (PNRR). FB acknowledges financial support from the TOPMASQ Project, CUP E13C24001560001, funded by the National Quantum Science and Technology Institute (NQSTI), PE0000023 of the PNRR, financed by the European Union – NextGenerationEU. RE acknowledges funding by the Deutsche Forschungsgemeinschaft (DFG, German Research Foundation) under Projektnummer 277101999 - TRR 183 (project A02) and under Germany's Excellence Strategy - Cluster of Excellence Matter and Light for Quantum Computing (ML4Q) EXC 2004/1 - 390534769.
	
	\appendix
	
	\section{Solution of axion electrodynamics}\label{app:solution}
	
	In this section, we provide some detail about the solution of the set of axion electrodynamics 
	equations (2)-(5) in cylindrical coordinates. 
	It is useful to massage the latter and bring it to the form
	\begin{equation}\label{delxB-N}
		\nabla\times\mathbf{B}=- i \frac{\omega}{c^2}\mathcal{E} \mathbf{E}-\frac{2\alpha}{\pi c}\mathbf{b}\times\mathbf{E} .
	\end{equation}
	Taking the curl of Eq. (3) and using Eq. (5), one obtains the wave equation
	\begin{equation}\label{eom}
		0=\nabla\left(\nabla\cdot\mathbf{E}\right)-\Delta\mathbf{E}-Q^2\mathbf{E}+
		i\omega\frac{2\alpha}{\pi c}\mathbf{b}\times\mathbf{E},
	\end{equation}
	where $Q^2= \frac{\omega^2\mathcal{E}(\omega)}{c^2}$.
	Choosing the $z$ axis along the main axis of the material, the wavevector component $q_z$ parallel 
	to the Weyl node separation vector $2\mathbf{b}$ is conserved.
	Moreover, the cylindrical symmetry implies that the electric field can be chosen as an eigenstate 
	of the generator of rotations around the $z$ axis
	\begin{equation}
		J_z=-i\partial_\varphi+\Sigma_z\;,\qquad \Sigma_z=\left(
		\begin{array}{ccc}
			0  & -i  &  0 \\
			i  &  0  &  0 \\
			0  &  0  &  1
		\end{array} \right)\;.
	\end{equation}
	It follows that it must have the dependence $\sim e^{im\varphi}$
	on the angular variable $\varphi$ and its components must take the form
	\begin{align}
		{E_r}(r)&=f(r) +g(r)\;, \nonumber\\
		{E_\varphi}(r)&=if(r) -i g(r)\;, \\
		{E_z}(r)&=h(r)\;, \nonumber
	\end{align}
	where $f(r), g(r), h(r)$ are complex functions of the radial coordinate $r$.
	Guided by the solution in the metallic limit $\mathbf{b}=0$, 
	one can make the Ansatz
	\begin{align}
		{E_r}&=A_m \mathcal{D}_{m-1}\left(q_\perp r\right)+B_m \mathcal{D}_{m+1}\left(q_\perp r\right)\;, \nonumber\\
		{E_\varphi}&=iA_m\mathcal{D}_{m-1}\left(q_\perp r\right)-iB_m \mathcal{D}_{m+1}\left(q_\perp r\right)\;, \\
		{E_z}&=C_m \mathcal{D}_{m}\left(q_\perp r\right)\;, \nonumber
	\end{align}
	where $\mathcal{D}_m$ denotes any Bessel function of the first kind of order $m$,
	$q_\perp$ is a parameter, in general complex, to be determined. In the bulk problem, $q_\perp$ is the
	modulus of the radial component of the momentum, $q_\perp=\sqrt{q_x^2+q_y^2}$.
	With the above Ansatz, the modified Maxwell equations yield algebraic 
	equations in the unknown coefficients $A_m$, $B_m$, $C_m$. In particular, Eq. (2) becomes
	\begin{equation}\label{GaussAlgebraic}
		-q_\perp\left(1-b_{\omega}\right)A_m+q_\perp\left(1+b_{\omega}\right)B_m+iq_z C_m =0 ,
	\end{equation}
	having defined the frequency-dependent dimensionless constant 
	$b_{\omega}={2\alpha b c}/{\pi\omega\mathcal{E}}$.
	We then write \eqref{eom} as the linear system
	\begin{widetext}
		\begin{equation}\label{coeffmat}
			\left(\begin{array}{ccc}
				Q^{2}-q^{2}-b_{\omega}\left(Q^{2}-\frac{q_\perp^{2}}{2}\right) & \frac{q_\perp^{2}}{2}b_{\omega} & 0\\
				-\frac{q_\perp^{2}}{2}b_{\omega} & Q^{2}-q^{2}+b_{\omega}\left(Q^{2}-\frac{q_\perp^{2}}{2}\right) & 0\\
				ib_{\omega}q_\perp q_z & ib_{\omega}q_\perp q_z & Q^{2}-q^{2}
			\end{array}\right)\left(\begin{array}{c}
				A_m\\
				B_m\\
				C_m
			\end{array}\right)	=	\left(\begin{array}{c}
				0\\
				0\\
				0
			\end{array}\right)   \;.
		\end{equation}
	\end{widetext}
	In order to find a nontrivial solution to \eqref{coeffmat}, the determinant of the matrix above must vanish. 
	This directly leads to the condition (9) in the main text, 
	which implicitly determines the dispersion relation of bulk modes \cite{Hofmann2016}.
	One finds two linearly independent electric field eigenmodes whose components are determined 
	from the solution of \eqref{GaussAlgebraic} and \eqref{coeffmat}.
	They can be written in the form
	\begin{equation}
		\mathbf{E}_m = a_m \left(\begin{array}{c}
			\frac{m}{q_\perp r}\mathcal{D}_m  \\
			i \mathcal{D}_m'\\
			0
		\end{array}\right)
		+b_m
		\left(\begin{array}{c}
			\mathcal{D}_m'  \\
			\frac{i m}{q_\perp r} \mathcal{D}_m\\
			-\frac{iq_zq_\perp}{Q^2-q_\perp^2}\mathcal{D}_m
		\end{array}\right)
	\end{equation}
	in which the coefficients $a_m=A_m+B_m$, $b_m=A_m-B_m$ are constrained by \eqref{GaussAlgebraic} and \eqref{coeffmat}.
	
	Analytic continuation $q_\perp \to -i\kappa$ allows us to describe modes which are exponentially localized at the WS-dielectric interface. For this class of solutions, the electric field is written as $\mathbf{E}_m=\sum_{s=\pm}a_s^{(m)}\mathbf{E}_{m,s}$, where the modes labeled by $s=\pm$ are 
	\begin{equation}\label{Ein}
		\mathbf{E}_{m,s}=\frac{\omega}{q_z}
		\left[\left(\begin{array}{c}
			\frac{im}{\kappa_s r}I_m\left(\kappa_sr\right)  \\
			- I_m'\left(\kappa_sr\right)\\
			0
		\end{array}\right)
		+\gamma_{s}
		\left(\begin{array}{c}
			iI_m' \left(\kappa_sr\right) \\
			-\frac{ m}{\kappa_s r} I_m\left(\kappa_s r\right)\\
			\frac{c\,q_z\kappa_s}{Q^2+\kappa_s^2}I_m\left(\kappa_sr\right)
		\end{array}\right)
		\right] \, ,
	\end{equation}
	in which the Bessel functions of the second kind $I_m$ are regular in the origin and
	\begin{equation}
		\gamma_{\pm}=
		\frac{1}{\overline{Q}}\left(\frac{\beta}
		{\sqrt{\mathcal{E}}}\pm\sqrt{\overline{q}_z^2+\left(\frac{\beta}{\sqrt{\mathcal{E}}}\right)^2}\right)\,.
	\end{equation}
	Here and in the following we denote the wavevectors in units of $\omega_p/c$ 
	as \mbox{$\overline{q}_z=q_zc/\omega_p$} and \mbox{$\overline{Q}=Qc/\omega_p$.}
	In the two solutions, the index $s$ appears in the argument of the Bessel functions and determines the localization length as discussed in the main text.
	The magnetic field  is readily computed via Eq. (3) in the frequency domain as
	\begin{equation}
		\mathbf{B}_{m,s}=
		\left(\begin{array}{c}
			I_m'\left(\kappa_sr\right) \\
			\frac{i\,m}{\kappa_sr}I_m\left(\kappa_sr\right)\\
			\frac{i\kappa_s}{q_z}I_m\left(\kappa_sr\right)
		\end{array}\right) 
		+\frac{1}{\gamma_s}
		\left(\begin{array}{c}
			\frac{i\,m}{\kappa_sr}I_m\left(\kappa_sr\right) \\
			iI_m'\left(\kappa_sr\right)\\
			0	\end{array}\right) .
	\end{equation}
	Because of the chosen parametrization, our solution contains a curve, identified by the condition 
	$\kappa_+=\kappa_-$, in which the two modes with $s=\pm$ are not linearly independent. 
	The degeneracy curve is defined from Eq. (10) by the condition 
	\begin{equation}\label{degeneratecondition}
		\overline{q}_z^2\mathcal{E}+\beta^2=0\;.
	\end{equation}
	Using the single-band approximation, i.e., retaining only the first term in the permittivity (6), one finds
	\begin{equation}\label{degenerate}
		\omega_{deg}\left(q_z\right)\approx\frac{\sqrt{\epsilon_W}\omega_p}{\sqrt{\epsilon_W+{\beta^2}/{ \overline{q}_z^2}}}\;.
	\end{equation}
	This is a good approximation as long as $\omega_{deg}\ll\omega_p$. 
	
	The field in the outer region is well know \cite{Stratton,Pfeiffer1974}. 
	Again, we focus on evanescent solutions in the radial direction and write it as the superposition 
	\mbox{$\mathbf{E}_{m}^{(out)} = b_{1}^{(m)}\mathbf{E}_{m,1}+b_{2}^{(m)}\mathbf{E}_{m,2}$},
	where
	\begin{align}\label{Eout}
		\mathbf{E}_{m,1} &=\frac{i\omega}{q_z}\left(\begin{array}{c}
			\frac{m}{\kappa_{d}r}K_{m}\left(\kappa_{d}r\right)\\
			iK_{m}'\left(\kappa_{d}r\right)\\
			0
		\end{array}\right) \;, \\
		\mathbf{E}_{m,2} &=\left(\begin{array}{c}
			K_{m}'\left(\kappa_{d}r\right)\\
			\frac{im}{\kappa_{d}r}K_{m}\left(\kappa_{d}r\right)\\
			\frac{i\kappa_d}{q_z} K_{m}\left(\kappa_{d}r\right)
		\end{array}\right)\;.
	\end{align}
	Here the functions $K_{m}$ are regular as $r\to\infty$ and, in fact, exponentially decaying, with inverse decay length $\kappa_d=\sqrt{q_z^2-Q_d^2}$ and $Q_d=\omega\sqrt{\epsilon_d}/c$. The corresponding magnetic field is computed by means of the usual Maxwell-Faraday equation as
	\begin{align}
		\mathbf{B}_{m,1} &=\left(\begin{array}{c}
			K_{m}'\left(\kappa_{d}r\right)\\
			\frac{im}{\kappa_{d}r} K_{m}\left(\kappa_{d}r\right) \\
			\frac{i\kappa_d}{q_z} K_{m}\left(\kappa_{d}r\right)
		\end{array}\right)
		\;, \\
		\mathbf{B}_{m,2} &= \frac{Q^2_d}{q_z^2}\left(\begin{array}{c}
			\frac{m}{\kappa_{d}r}K_{m}\left(\kappa_{d}r\right)\\
			iK_{m}'\left(\kappa_{d}r\right)\\
			0
		\end{array}\right)\;.
	\end{align}

	\section{Boundary conditions for the metal-dielectric interface}\label{app:bc}
	\subsection{Derivation of the boundary conditions}
	We now provide some details about the manipulations of the modified electrodynamics equations. 
	The modified Amp\`ere-Maxwell law follows
	\begin{equation}\label{eq:velxB-G}
		\nabla\times\mathbf{B}= -i\frac{\omega}{c^{2}}\mathbf{E}+\mu_{0}\mathbf{j}_e-\frac{2\alpha}{\pi c}\mathbf{b}\times\mathbf{E}\;.
	\end{equation}
	It is customary to divide the current in the source term into the contributions from free and bound charges
	\begin{equation}
		\mathbf{j}_e	=	\mathbf{j}_{f}+\mathbf{j}_{p}=\sigma\mathbf{E} -i\omega \mathbf{P} .
	\end{equation}
	As we are considering a magnetic WS, no magnetization current is present in this expression. 
	In general, it can be present in the dielectric and handled in a standard way.
	For a linear material the polarization is proportional to the applied field $\mathbf{P}=\varepsilon_0\chi_e\mathbf{E}$ and \eqref{eq:velxB-G} becomes
	\begin{equation}\label{velxB-GN}
		\nabla\times\mathbf{B}=-i \frac{\omega}{c^2}\mathcal{E}\mathbf{E}-\frac{2\alpha}{\pi c}\mathbf{b}\times\mathbf{E}
	\end{equation}
	with
	$
	\varepsilon_{0}\mathcal{E}	=	\varepsilon_{0} \epsilon_W+i{\sigma}/{\omega}.
	$
	
	Let us consider an interface at $r=R$. Following the viewpoint of \cite{Langreth1989}, we derive the boundary condition on the macroscopic fields to lowest order in a gradient expansion. The localized contributions to charge and current density have been considered in \cite{Chen2019} and found to be not relevant for this class of problems in the small wavevector regime.
	Integrating Eq. \eqref{velxB-GN} across the interface, one establishes the continuity of the components of the magnetic field. 
	This does not exclude a static uniform magnetization of the material, see \cite{LandauE}. We note that the contribution of the magnetization current is neglected throughout the paper $\nabla\times \mathbf{M}\approx0$.
	In symbols
	\begin{equation}
		\mathbf{B}_{\parallel}\left(r=R^+\right)=\mathbf{B}_{\parallel}\left(r=R^-\right)
	\end{equation}
	We approximate the corresponding magnetic constant as $\mu_d\approx\mu_0$. The above equation is therefore equivalent to the continuity of the magnetic field parallel to the interface, while straightforward modifications are necessary in the most general case.
	Using $\nabla\cdot\mathbf{P}=-\rho_{b}$, with $\rho_{b}$ the density of bound charge, one writes the total charge density as the sum of free and bound contributions
	\begin{equation}
		\rho_{e}	=	\rho_{b}+\rho_{f}=
		-\nabla\cdot\left(\varepsilon_{0}\chi_{e}\mathbf{E}+\frac{i\sigma}{\omega}\mathbf{E}\right)\;.
	\end{equation}
	Defining the electric displacement in the WS as
	\begin{equation}\label{del.D}
		\mathbf{D}=\left(\varepsilon_0\mathcal{E}\mathbf{E}+\frac{2\alpha c\varepsilon_0}{\pi i \omega}
		\mathbf{b}\times\mathbf{E}\right)
	\end{equation}
	and in the dielectric as $\mathbf{D}=\epsilon_d\varepsilon_0\mathbf{E}$,  
	Eq. (2) can be cast in the form
	\begin{equation}
		\nabla\cdot\mathbf{D}=0 \;.
	\end{equation}
	Integrating the previous equation across the interface, one obtains the continuity of the component perpendicular to the interface
	\begin{equation}\label{eldisp}
		\mathbf{D}_{\perp}\left(r=R^+\right)=\mathbf{D}_{\perp}\left(r=R^-\right).
	\end{equation}
	Integration of the homogeneous Maxwell equations (3) and (4) across the interface 
	determines the continuity of the  components of the electric field parallel to the interface
	\begin{equation}
		\mathbf{E}_{\parallel}\left(r=R^+\right)=\mathbf{E}_{\parallel}\left(r=R^-\right)
	\end{equation}
	and of the magnetic field perpendicular to it
	\begin{equation}
		\mathbf{B}_{\perp}\left(r=R^+\right)=\mathbf{B}_{\perp}\left(r=R^-\right)\;.
	\end{equation}

	\subsection{Imposing the boundary conditions}
	Let us now consider the WS-dielectric interface at $r=R$. We have established that the electric field inside the semimetallic cylinder is a linear superposition of the modes \eqref{Ein}, with coefficients $a_\pm^{(m)}$, while in the dielectric outside it is a linear superposition of the modes \eqref{Eout}, with coefficients $b_{1,2}$.
	
	As the Maxwell equations completely determine the magnetic field once the electric field is known, three conditions are needed at the dielectric-WS interface.
	A more compact form is obtained by instead imposing four conditions and requiring that they are compatible among them \cite{Stratton}. The continuity of the tangential components of the electric field and of the magnetic induction, in particular, results in a linear system for the four unknown coefficients  $\mathbf{c}_m=\left(a_+^{(m)},a_-^{(m)},b_{1}^{(m)},b_{2}^{(m)}\right)$. After an appropriate rescaling of the coefficients, the system can be cast in the form $B_m \mathbf{c}_m=0$, with the matrix
	\begin{widetext}
		\begin{equation} \label{BCmatrix}
			B_m  = 
			\left(\begin{array}{cccc}
				\frac{I'_{m}\left(u_{+}\right)}{I_{m}\left(u_{+}\right)}+\frac{\gamma_{+}m}{u_{+}} & \frac{I'_{m}\left(u_{-}\right)}{I_{m}\left(u_{-}\right)}+\frac{\gamma_{-}m}{u_{-}} & \frac{K'_{m}\left(v\right)}{K_{m}\left(v\right)} & \frac{m}{v}\\
				\frac{u_{+}}{\gamma_{+}}\frac{q_{z}^{2}}{Q^{2}} & \frac{u_{-}}{\gamma_{-}}\frac{q_{z}^{2}}{Q^{2}} & 0 & v\\
				\frac{I'_{m}\left(u_{+}\right)}{\gamma_{+}I_{m}\left(u_{+}\right)}+\frac{m}{u_{+}} & \frac{I'_{m}\left(u_{-}\right)}{\gamma_{-}I_{m}\left(u_{-}\right)}+\frac{m}{u_{-}} & \frac{m}{v} & \frac{Q_{d}^{2}}{q_{z}^{2}}\frac{K'_{m}\left(v\right)}{K_{m}\left(v\right)}\\
				u_{+} & u_{-} & v & 0
			\end{array}
			\right)\; .
		\end{equation}
	\end{widetext}
	The arguments of the Bessel functions are denoted as
	\begin{equation}
		u_{\pm}	=\kappa_{\pm}R=\rho\sqrt{\overline{q}_z^{2}-\overline{Q}^{2}
			+\frac{2\beta^{2}}{\mathcal{E}}
			\pm\frac{2\beta }{\sqrt{\mathcal{E}}}\sqrt{\overline{q}_z^{2}+\frac{\beta^{2}}{\mathcal{E}}}}\; \label{upmdef}
	\end{equation}
	and
	\begin{equation}
		v=\kappa_{d}R\,=\,\rho \sqrt{\overline{q}_z^{2}-\overline{Q}_{d}^{2}} \;. \label{vdef}
	\end{equation}
	The boundary conditions can only be consistently satisfied if the determinant of this matrix vanishes, which amounts to the condition
	\begin{equation}\label{detBm=0}
		\det B_m\left(\omega,q_z\right)\,=\,0\;.
	\end{equation}
	This equation implicitly determines the dispersion curves $\omega=\omega_m(q_z)$. As for the metallic waveguide, it must be solved numerically in the general case. In the next section, we provide useful starting points for the root-finding routines. The solutions with $-10\le m\le 10$ are illustrated for sample parameters in Fig. \ref{fig:full}.
	
	We note that the SPPs obtained by solving Eq.~\eqref{detBm=0} propagate along
	the direction of the separation between the Weyl nodes. This situation corresponds to a Faraday 
	configuration, where, in a planar slab, the modes are reciprocal \cite{Kotov2018}. In our case, 
	the dispersions remain symmetric under inversion of $q_z$ but become non reciprocal with respect
	to the angular momentum $m$.
	
	\section{Asymptotic regimes}\label{app:asymptotes}
	
	\subsubsection*{Metallic limit}\label{app:b->0}
	We check now that the limit $b\to0$ (or, equivalently, $\beta\to0$) correctly reproduces the plasmonic dispersion of the metallic waveguide. In such limit, the arguments of the Bessel functions in \eqref{upmdef} tend to the same expression \mbox{$u_\pm\to u=R\sqrt{q_z^2-Q^2}$}, which is the corresponding value in a metal \cite{Stratton}. In the same limit \mbox{$\gamma_\pm\to\pm\gamma$}, where $\gamma= \sqrt{{q_z^2}/{Q^2}}$. The determinant of \eqref{BCmatrix} is most easily computed by replacing the first two columns by linear combinations, namely, semi-sum and semi-difference, so that one arrives at the matrix
	\begin{equation}\label{BCb->0}
		B_m^{(0)}= 		\left(\begin{array}{cccc}
			\frac{I'_{m}\left(u\right)}{I_{m}\left(u\right)} & \frac{\gamma m}{u} & \frac{K'_{m}\left(v\right)}{K_{m}\left(v\right)} & \frac{m}{v}\\
			0 & \frac{u}{\gamma}\frac{q_{z}^{2}}{Q^{2}} & 0 & v\\
			\frac{m}{u} & \frac{I'_{m}\left(u\right)}{\gamma I_{m}\left(u\right)} & \frac{m}{v} & \frac{Q_{d}^{2}}{q_{z}^{2}}\frac{K'_{m}\left(v\right)}{K_{m}\left(v\right)}\\
			u & 0 & v & 0
		\end{array}
		\right) \;.
	\end{equation}
	The condition $\det B_m^{(0)}=0$ is exacly the one that implicitly defines the frequencies of the SPP modes in a metal \cite{Pfeiffer1974}. 
	
	\subsubsection*{Planar limit}\label{app:planar}
	
	In order to make contact with known results \cite{Hofmann2016,Kotov2016,Kotov2018}, 
	we now discuss the planar limit of the cylindrical waveguide. We set the wavevector in the angular direction $q_y=m/R$ 
	and take the limit $R\to\infty$ and $|m|\to\infty$, while keeping their ratio $q_y$ fixed.
	Then the factor $e^{im\varphi}$ directly maps into the plane wave $e^{iq_yy}$, with $y=R\varphi$.
	The radial and azimuthal versors $\hat r$ and $\hat \varphi$ are mapped into the versors $\hat x$ and $\hat y$ respectively. 
	Next, we need to calculate the limit of the ratios
	\begin{equation}
		\frac{I'_m(\kappa_\pm R)}{I_m(\kappa_\pm R)}, \quad \frac{K'_m(\kappa_d R)}{K_m(\kappa_d R)}.
	\end{equation}
	We focus on the first ratio above, for the second the calculation is essentially the same and we only 
	give the final result.
	Omitting for notational simplicity the index $\pm$, we set 
	$\nu = |m| = R|q_y|$, $t = \kappa/|q_y|$
	and write 
	\begin{align}
		\frac{I'_{|m|}(\kappa R)}{I_{|m|}(\kappa R)} = \frac{I'_{\nu}(\nu t)}{I_{\nu}(\nu t)}= \frac{1}{\nu} \frac{d}{dt}\ln I_\nu(\nu t).
	\end{align}
	We also define
	$$
	\mu_\pm = \sqrt{q_y^2+\kappa_\pm^2}, \quad \mu_d =  \sqrt{q_y^2+\kappa_d^2}.
	$$
	which are inverse decay lengths in the $x$-direction, corresponding to 
	the evanescent waves $e^{\mu_\pm x}$ and $e^{-\mu_d x}$ of the planar geometry. 
	We recall the homogeneous expansions of the modified Bessel functions \cite{NISTDLMF}
	\begin{align}\label{Ihomo}
		I_\nu(\nu t) &\sim \frac{e^{\nu \eta_t}}{(2\pi \nu)^{1/2}(1+t^2)^{1/4}} , \\
		K_\nu(\nu t) & \sim  \frac{\pi e^{-\nu \eta_t}}{(2\pi \nu)^{1/2}(1+t^2)^{1/4}} ,
	\end{align}
	where 
	\begin{align}
		\eta_t & = \sqrt{1+t^2} + \ln \frac{t}{1+\sqrt{1+t^2} },\\
		p_t & = \frac{1}{\sqrt{1+t^2}}.
	\end{align}
	These expansions hold uniformly for $0<t<\infty$ as \mbox{$\nu \to + \infty$}. 
	Using \eqref{Ihomo}, we obtain
	\begin{align}
		\frac{I'_m(\kappa R)}{I_m(\kappa R)} &\sim \frac{d\eta_t}{dt} = \frac{\sqrt{q_y^2+\kappa^2}}{\kappa} ,
		\\
		\frac{K'_m(\kappa_d R)}{K_m(\kappa_d R)} &\sim  - \frac{d\eta_t}{dt} = - \frac{\sqrt{q_y^2+\kappa_d^2}}{\kappa_d} \;.
	\end{align}
	Restoring the label $s=\pm$, the matrix encoding the interface conditions becomes
	\begin{equation}
		B_m \rightarrow 
		\begin{pmatrix}
			\mu_+  + \gamma_+ q_y  & 
			\mu_-  + \gamma_- q_y&
			- \mu_d &  q_y\\
			\frac{\left( - \mu_+^2 + q_y^2 \right)q^2_z}{\gamma_+ Q^2} &  
			\frac{\left(  -\mu_-^2 + q_y^2 \right)q^2_z}{\gamma_-Q^2} & 0 &  - \mu_d^2 + q_y^2   \\
			q_y  +  \frac{\mu_+}{\gamma_+} &  q_y
			+  \frac{\mu_-}{\gamma_-}  & q_y   &  
			- \frac{Q_d^2}{q_z^2}     \mu_d  \\
			- \mu_+^2 + q_y^2 & - \mu_-^2 + q_y^2 & - \mu_d^2 + q_y^2 & 0 
		\end{pmatrix} , \label{limitM}
	\end{equation}    
	This matrix coincides with the one obtained for the half-space geometry, 
	with the WS with ${\bf b}=b\hat z$ in the region $x<0$ 
	and the dielectric in the region $x>0$ \cite{Hofmann2016}.

	\begin{figure}[h]
		\centering
		\includegraphics[width=\linewidth]{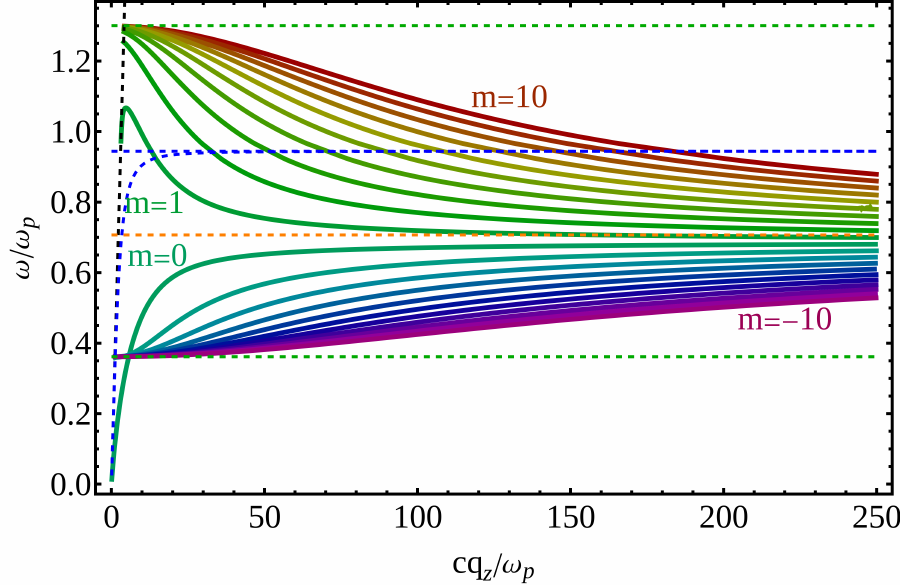}
		\caption{Crossover in the SPP dispersions for $\beta=10$,
			$\rho=0.1$, $\omega_c/\omega_p=10$, $\omega_F/\omega_p=1$. The vertical order of the negative-$m$ solutions as $q_z\to\infty$ is inverted with respect to the order at $q_z\to0$. The green dashed lines are the $m\to\pm\infty$ asymptotes from \eqref{masymptotecondition}, the blue dashed lines are the plasma frequency computed from the full permittivity (6)
			and the degeneracy line $\kappa_+=\kappa_-$, while the orange dashed line is the large-wavevector limit computed from \eqref{largeqzdet}.}
		\label{fig:full}
	\end{figure}
	
	\subsubsection*{Limit of large wavevector}
	We provide here some detail about the large-wavevector limit (12). 
	We make use of the asymptotic expansions of the Bessel functions of the second kind for large argument $z$ \cite{NISTDLMF}
	\begin{align}
		\frac{I_{m}'\left(z\right)}{I_{m}\left(z\right)}&\sim 1-\frac{1}{2z}, 
		\\
		\frac{K_{m}'\left(z\right)}{K_{m}\left(z\right)}&\sim -1-\frac{1}{2z}. 
	\end{align}
	Expanding \eqref{upmdef} and \eqref{vdef}, one arrives at the expression
	\begin{equation} \label{largeqzdet}
		\det M_{m}	\sim -4q_{z}R\left(Q^{2}+Q_{d}^{2}\right)+2\left(Q^{2}-Q_{d}^2\right)+8Q\frac{m\alpha b}{\pi\sqrt{\mathcal{E}}}\;.
	\end{equation}
	Substituting the definitions of $Q$ and $Q_d$ and solving the vanishing condition of the determinant in $\omega$ with the high-frequency, single-band approximation for the dielectric function, one finds Eq. (12). 
	The asmptotic limit as computed from Eq. \eqref{largeqzdet} is shown as an orange dashed line in Fig. \eqref{fig:full}.
	
	\subsubsection*{Limit of large angular momentum}
	We consider now the limit $m\to\pm\infty$ at fixed wavevector. We need the identities
	\begin{equation}
		\frac{I_m'(u)}{I_m(u)}\sim \frac{|m|}{u}\quad,\qquad \frac{K_m'(v)}{K_m(v)}\sim -\frac{|m|}{v}\;,
	\end{equation}
	for the order of the Bessel functions $m\to\pm\infty$. After some algebra, one obtains Eq. \eqref{detBm=0} in the form
	\begin{equation}
		\left[\overline{\omega}\left(\epsilon_d+\mathcal{E}\right)-2\mbox{sign}(m)\beta\right]
		\left[\overline{q}_z^2-\overline{\omega}^2\mathcal{E}-2\mbox{sign}(m)\beta\overline{\omega}\right]=0 \,,
	\end{equation}
	apart from an overall factor, different from zero. (We use the notation $\overline{\omega}=\omega/\omega_p$.)
	Requiring that the solutions  have a non-diverging localization length and reproduce the results for the metallic cylinder in the limit $\beta \to0$ implies that the $m\to\pm\infty$ asymptotes must satisfy
	\begin{equation}\label{masymptotecondition}
		\overline{\omega}\left(\epsilon_d+\mathcal{E}\right)-2\mbox{sign}(m)\beta=0\;.
	\end{equation}
	In general, this has to be solved numerically, but a good approximation is obtained when retaining only the first term in Eq. (6), 
	which yields the $\omega_\pm$ in Eq. (13) of the main text. The numerical solutions of Eq. \eqref{masymptotecondition} are instead shown in Fig. \ref{fig:full} as green dashed lines.

	\section{Parameters}\label{app:estimates}
	As an example, we consider the set of parameters \mbox{$E_F\approx 0.04\,$eV}, $v_F=10^5\,$m/s, $\omega_F \approx 6.1\times10^{13}\,$Hz, 
	\mbox{$E_c\approx 0.2\,$eV}, $\omega_p\approx 5.8\times10^{13}\,$Hz, modeled on
	$\mbox{Eu}\mbox{Cd}_{2}\mbox{As}_{2}$ \cite{Wang2019,Krishna2018,Wang2016}, see also \cite{Bradlyn2017,Vergniory2019,SantosCottin2023}. 
	One finds $\omega_c/\omega_{p}\sim5$, $\omega_F/\omega_{p}\sim1$ and $\beta\sim7$. The  parameter $\rho=0.1$ used in the main text corresponds 
	to a cylinder of $R\approx0.5\,\mu$m. The high-frequency skin depth is then estimated as $\delta_0\sim 5\,\mu$m and reference wavevector $\omega_p/c\sim 2\times10^5\mbox{m}^{-1}$. 
	The minimal wavelength in Fig. 3 is therefore in the visible range.

	\bibliography{PRRsubmission}
	
\end{document}